\documentclass[aps,pra,showpacs,twocolumn]{revtex4}

\usepackage{bm}
\usepackage{graphicx}
\usepackage{amsmath}
\usepackage{latexsym}
\usepackage{amsfonts}
\usepackage{amssymb}
\usepackage{array}

\newcommand{\ket}[1]{\left\vert#1\right\rangle}
\newcommand{\miniket}[1]{\vert#1\rangle}

\newcommand{\bra}[1]{\left\langle#1\right\vert}
\newcommand{\minibra}[1]{\langle#1\vert}

\begin{document}

\title{Conditions for Factorizable Output From a Beam splitter}

\author{S. C. Springer,$^1$ Jinhyoung Lee,$^2$ M. Bellini,$^{3,4}$ and M. S. Kim$^1$}

\affiliation{$^1$ School of Mathematics and Physics, Queen's University, Belfast BT7 1NN, United Kingdom}
\affiliation{$^2$ Department of Physics, Hanyang University, Seoul, 133-791, Korea}
\affiliation{$^3$ European Laboratory for Nonlinear Spectroscopy (LENS), Via Nello Carrara 1, 50019 Sesto Fiorentino, Florence, Italy}
\affiliation{$^4$ Instituto Nazionale di Ottica Applicata (CNR), Largo E. Fermi, 6, I-50125, Florence, Italy.}

\date{\today}

\begin{abstract}
A beam splitter is one of the most important devices in an optics laboratory because of its handiness and versatility; equivalent devices are found in various quantum systems to couple two subsystems or to interfere them.  While it is normal that two independent input fields are superposed at the beam splitter to give correlated outputs, identical Gaussian states interfere there to produce totally independent output fields.  We prove that Gaussian states with same the variance are the only states which bring about factorizable output fields.
\end{abstract}


\maketitle

\section{Introduction}
Quantum state preparation is the manipulation of a certain state into a desired target state. It is crucial for both the testing and implementation of quantum mechanics  to be able to generate custom states (or custom states within a specified tolerance)~\cite{generalQSE}.

There are two main approaches to quantum state preparation: either by the time evolution of the state using a Hamiltonian that transforms the state from its initial condition to the final target state or by conditional preparation. In the second approach a measurement is made on one part of a bipartite correlated system; the action of this measurement is to modify the correlated system with the aim of producing a useful state. Paris {\em et al}. discuss the conditional preparation of a state in one mode of a two mode squeezed state by an appropriate measurement on the other mode in  Ref. \cite{ParisConditionalMeasurement}. Conditional preparation is an essential feature of teleportation protocols \cite{teleportation,OpatrnyEtAl}.

An important application of the conditional preparation method is the Knill-Laflamme-Milburn (KLM) scheme \cite{KLM} for optical quantum computing based on the use of measurements to evolve states by exploiting symmetry relations satisfied by bosons. This work was seminal as it theoretically demonstrated the possibility of scalable linear optical computing. Previous implementations of quantum circuits had been proposed which were linear, Cerf {\em et al.} \cite{Cerf} had put forward an all linear scheme that avoided introducing the non-linear Kerr effects which were a common feature of schemes up until then (for example \cite{Milburn}). The KLM scheme does not suffer from the scalability problems of the other proposals which required an exponential increase in the number of optical elements when processing increasing numbers of qubits. Much work has been done since, to refine this protocol and improve its efficiency \cite{Yoran}.

For continuous-variable fields, it was proven~\cite{dakna} that a quantum state can be engineered by single photon additions together with displacements~\cite{paris-d}.  As the photon addition is relatively difficult~\cite{bellini}, Fiur\'{a}\v{s}ek {\em et al.}. \cite{fiu} came up with an idea to subtract photons from a squeezed field to generate a desired quantum state.  Photon subtraction has been realized in various laboratories using a beam splitter and a single photon detector~\cite{sub}.

In quantum state engineering, two output fields are in general (quantum-mechanically or classically) correlated.  In this way, by measuring one field, the other field is collapsed into a desired state. It was shown that when nothing is injected into one input port of a beam
splitter, the other input has to be a coherent state for the two
output fields to be factorizable \cite{Zavatta}. However, we can question if there are general input states which give totally uncorrelated outputs.  In this paper, we seek to answer this question.

We begin with a brief review of the topic of beam splitters (Sec.~\ref{Sec:Beamsplitters}) and the ability to perform conditional preparation with only classically correlated states (Sec.~\ref{Sec:CondPrepCCstate}). In Sec.~\ref{Sec:Correlations} we examine the conditions required for the output of a beam splitter to be factorizable, and hence not correlated. The result that factorizable output requires Gaussian states as input is presented and this leads us to specify that, in fact, the Gaussian states may differ only by an arbitrary displacement.
An alternative description, in the Wigner function representation, is also examined.

\section{State preparation with a beam splitter}
\label{Sec:Beamsplitters}
Beam splitters are ubiquitous in quantum optics and are an essential component of almost  all photonic quantum state engineering protocols. They are passive (energy conserving) devices which are composed of a linear medium; that is the polarisation is proportional to the incoming field with constant of proportionality given by the first order susceptibility $\chi^{(1)}$. They may be two attatched glass prisms or a partially silvered mirror and are constructed to a given specification so as to transmit a certain proportion of an incoming source and reflect a corresponding amount as quantified by the coefficients of transmittivity and reflectivity ($t=\cos{\frac{\theta}{2}}$ and $r=\sin{\frac{\theta}{2}}$ respectively and satisfying $r^2 + t^2 =1$).  For two incident beams the beam splitter facilitates interference of the two input fields to create two, generally different, output fields.

Quantum mechanically we describe the action of a beam splitter by the operation~\cite{campos}
\begin{eqnarray}
\hat{B}(\theta) = \exp [i {\frac{\theta}{2}} (\hat{a}^\dag \hat{b} + \hat{a} \hat{b}^{\dag})] \label{beamsplitter}
\end{eqnarray}
where $\hat{a}$ and $\hat{b}$ are the annihilation operators of the two input modes and $\hat{a}^\dag$ and $\hat{b}^\dag$ are their Hermitian conjugates.
The need for the device to preserve the bosonic commutation relations in the output modes prescribes that the operator must be unitary. The trivial cases of $r=0$ or $t=0$ are ignored since in these situations the beam splitter is not mixing the two input modes and so no correlations may arise.

In the Heisenberg picture the action of the beam spitter is to effect the following changes:
\begin{equation}
\label{eq:actionbs}
\hat{a} \to  t \hat{a} + r \hat{b} ~~~,~~~\hat{b} \to t \hat{b} - r \hat{a}
\end{equation}
Defining quadrature operators $\hat{q}_a = \frac{1}{\sqrt{2}} (\hat{a}^{\dag} + \hat{a})$, $\hat{p}_a = \frac{i}{\sqrt{2}} (\hat{a}^{\dag} - \hat{a})$, and similarly $\hat{q}_b$ and $\hat{p}_b$,
the relations describing the first order transformations (\ref{eq:actionbs}) may be rewritten in terms of the quadratures $q_i$ and $p_i$ as
\begin{eqnarray}
 \begin{pmatrix}
 \hat{q}_a \\
 \hat{p}_a \\
 \hat{q}_b \\
 \hat{p}_b
 \end{pmatrix}
 \to
 \begin{pmatrix}
 t & 0 & r & 0 \\
 0 & t & 0 & r \\
 -r & 0 & t & 0 \\
 0 & -r & 0 & t
 \end{pmatrix}
\begin{pmatrix}
 \hat{q}_a \\
 \hat{p}_a \\
 \hat{q}_b \\
 \hat{p}_b
 \end{pmatrix}.
 \label{firstorderBS}
\end{eqnarray}

\subsection{State Modification by Measurement of a Classically Correlated State} \label{Sec:CondPrepCCstate}
To implement conditional preparation requires two correlated states, however this correlation need not be entanglement---classical correlations can be suffcient. Consider the case of two distinct thermal states which represent equilibrium states for given temperatures $T_i$ ($i=a,b$):
\begin{eqnarray}
\hat{\rho}_i &=& \frac{1}{\pi \, \bar{n}_i} \int \exp  \left( \frac{-|\alpha|^2}{\bar{n}_i} \right) \miniket{\alpha}_{ii} \minibra{\alpha} \, d^2\alpha,
\end{eqnarray}
where $\bar{n}_i$ is the mean number of photons and $\ket{\alpha}$ represents a coherent state of amplitude $\alpha$.
Impinging two distinct thermal fields $\hat{\rho}_a$ and $\hat{\rho}_b$ onto a beam splitter generates the state with density operator
\begin{eqnarray}
\hat{\rho}_{ab}=\frac{1}{\pi^2 \bar{n}_a \bar{n}_b} \int \exp  \left( -\frac{|t \alpha- r \beta|^2}{\bar{n}_a} -\frac{|r \alpha + t \beta|^2}{\bar{n}_b} \right) \times \nonumber \\
(\miniket{\alpha}_{aa}\minibra{\alpha} \otimes \miniket{\beta}_{bb}\minibra{\beta}) \, d^2 \alpha \, d^2 \beta. \label{BStwothermals}
\end{eqnarray}
Using the definition of the two-mode Glauber-Sudarshan $P$-function ($P$ function for simplicity)~\cite{IntroQOGerryKnight}
\begin{eqnarray}
\hat{\rho}= \int P(\alpha, \beta) (\miniket{\alpha}_{aa} \minibra{\alpha} \otimes \miniket{\beta}_{bb} \minibra{\beta}) \, d^2 \alpha \, d^2 \beta
\end{eqnarray}
and Eq.~(\ref{BStwothermals}) we easily find the $P$ function for the output field to be
\begin{eqnarray}
P(\alpha, \beta) = \frac{1}{\pi^2 \bar{n}_a \bar{n}_b} \exp  \left( -\frac{|t \alpha- r \beta|^2}{\bar{n}_a} -\frac{|r \alpha + t \beta|^2}{\bar{n}_b} \right)
\end{eqnarray}
This $P$ function is a smooth, non-negative function and thus the state (\ref{BStwothermals}) describes a state which is classically correlated~\cite{kim}.   Despite the lack of quantum correlations this classical state may be used in an implementation of conditional measurement. For example, if a measurement results in the second mode being projected onto the one photon Fock state then mode $a$ becomes the non-Gaussian state
\begin{eqnarray}
\frac{1}{\pi A \, \bar{n}_a \bar{n}_b}  \int \exp \left[ -\left( \frac{t^2}{\bar{n}_a} + \frac{r^2}{\bar{n}_b} - \frac{B^2}{A} \right) |\alpha|^2 \right] \times \nonumber \\
\left( \frac{B^2}{A^2} |\alpha|^2 + \frac{1}{A} \right) \miniket{\alpha}_{aa} \minibra{\alpha} \, d^2 \alpha
\end{eqnarray}
where $A=1+\frac{r^2}{\bar{n}_a} + \frac{t^2}{\bar{n}_b}$ and $B=rt \left( \frac{1}{\bar{n}_b} - \frac{1}{\bar{n}_a}\right)$. By measurements on one mode of a classically correlated bipartite state a non-Gaussian state has been engineered in the other mode.
Thus, even classically correlated states may be useful in quantum state engineering by the method of conditional measurement. Here we have produced a non-classical state \cite{petr}

\section{Factorizable Output from a Beam Splitter}
\label{Sec:Correlations}

In this section, we find the condition for factorizable input fields of a beam splitter to result in completely independent output fields. Such states can be of no use in quantum state preparation. 

A density operator, $\hat{\rho}$, for continuous-variable fields is assumed to be written as the exponential of a Hermitian operator $\hat{f}$ as
\begin{eqnarray}
\hat{\rho}= e^{\hat{f}(\hat{a},\hat{a}^\dag)}. \label{expoenentialformofrho}
\end{eqnarray}
Continuous-variable fields necessarily involve unbounded operators such as the quadratures $\hat{q}$ and $\hat{p}$. Such unbounded operators have spectra that may go to infinity and this fact implies that unbounded operators should be definied on certain subspaces of the Hilbert space due to the Hellinger-Toeplitz theorem \cite{ReedSimonFuncAnalysis}. For these cases, every functional or density operator can be defined by employing unbounded operators such as $\hat{a}$ and $\hat{a}^\dag$ on their intersected subspace from spectral theory \cite{Peresbook}. The density operator of Fock states and superpositions of Fock states can be written in the form~(\ref{expoenentialformofrho}).
This is not the case for all operators on a finite dimensional Hilbert space: A finite-dimensional density operators may have zero eigenvalue(s) and so may not be written in such an exponential form with the exponent being any bounded operator; examples would be density operators corresponding to spin states, polarisation states and angular momentum states.

Equally we may define the functionals $\hat{\rho}$ and $\hat{f}$ in terms of the quadratures $\hat{q}$ and $\hat{p}$ as
\begin{eqnarray}
\label{ExpOperator}
\hat{\rho}= e^{\hat{f}(\hat{q}, \hat{p})}.
\end{eqnarray}
We shall represent the exponent $\hat{f}$ by a power series with respect to $\hat{q}$ and $\hat{p}$. The expansion could be made with respect to any ordering. As the exponent $\hat{f}$ is Hermitian, we choose symmetric ordering, denoted by  $\{ \cdot \}_s$; for example, $ \{ \hat{\xi}_i \hat{\xi}_j \}_s = \frac{1}{2!} ( \hat{\xi}_i \hat{\xi}_j + \hat{\xi}_j \hat{\xi}_i)$ and $\{ \hat{\xi}_i \, \hat{\xi}_j \, \hat{\xi}_k\}_s = \frac{1}{3!} (\mbox{sum of all permutations of } \hat{\xi}_i \hat{\xi}_j \hat{\xi}_k)$, where $\vec{\xi} = (\hat{q}, \hat{p})$. Then,
\begin{eqnarray}
\hat{f} = f_{(0)} + f_{(1)}^i \hat{\xi}_i + f_{(2)}^{ij} \{ \hat{\xi}_i \, \hat{\xi}_j \}_s +
f_{(3)}^{ijk} \{ \hat{\xi}_i \, \hat{\xi}_j \, \hat{\xi}_k\}_s + \cdots \nonumber \\ 
\label{polyform}
\end{eqnarray}
where $i,j,k,\ldots=1,2$. Here we employ Einstein's convention for the summations. Because of the ordering used, the $n$-th order coefficient tensor $f_{(n)}$ is of all permutations consisting of $n$-th moments in $\hat{q}$ and $\hat{p}$.
The fact that the exponent $\hat{f}$ is Hermitian and the expansion is made in the symmetric ordering in terms of $\hat{q}$ and $\hat{p}$ implies that all the coefficient tensors $f_{(n)}$ are real and symmetric. The coefficient tensors $f_{(n)}$ completely specify the state. The representation in Eq.~(\ref{polyform}) contains redundancies in the sense that all components of $f_{(n)}$ are not independent as $f_{(n)}$ is symmetric; for example, $f_{(2)}^{12}=f_{(2)}^{21}$. Remembering this redundance, we shall still employ the expansion in the form of Eq.~(\ref{polyform}) for mathematical convenience.

The transformation rules for $f_{(n)}$ follow once those for $\hat{\xi}_i$ are known. Supposing that $\hat{\xi}_i$ is transformed by applying a unitary operation as
\begin{eqnarray}
\hat{\xi}_i \to \hat{U} \hat{\xi}_i \hat{U}^\dag = \Lambda_i^j \, \hat{\xi}_j  \label{relation},
\end{eqnarray}
we then have the transformation rules for the (symmetric) higher moments, for example,
\begin{eqnarray}
\hat{U} \{\hat{\xi}_i \, \hat{\xi}_j \}_s \hat{U}^\dag &=& \Lambda_i^k \, \Lambda_j^l \, \{ \hat{\xi}_k \, \hat{\xi}_l \}_s, \nonumber \\
\hat{U} \{\hat{\xi}_i \, \hat{\xi}_j \, \hat{\xi}_k \}_s \hat{U}^\dag &=& \Lambda_i^l \, \Lambda_j^m \,  \Lambda_k^n \, \{ \hat{\xi}_l \, \hat{\xi}_m \, \hat{\xi}_n \}_s.
\end{eqnarray}
The transformed operators, $\hat{U} \hat{\xi}_i \hat{U}^\dag$, preserve the commutation relations of $\hat{\xi}_i$.
The transformed density operator obtained by applying the unitary operation is now represented as
\begin{eqnarray}
\hat{U} \hat{\rho} \hat{U}^\dag = \exp (\hat{U} \hat{f} \hat{U}^\dag).
\end{eqnarray}
Here, the transformed exponent is expanded as
\begin{eqnarray}
\hat{U} \hat{f} \hat{U}^\dag &=& \bar{f}_{(0)} + \bar{f}_{(1)}^{i} \, \hat{\xi}_i + \bar{f}_{(2)}^{ij} \, \{ \hat{\xi}_i \, \hat{\xi}_j \}_s + \ldots,
\end{eqnarray}
where the transformed coefficient tensors $\bar{f}_{(n)}$ are given in terms of $f_{(n)}$ and $\Lambda_i^j$ as
\begin{eqnarray}
\label{eq:tct1}
\bar{f}_{(0)} &=& f_{(0)}, \\
\label{eq:tct2}
\bar{f}_{(1)}^i &=& \Lambda_{j}^{i} f_{(1)}^j, \\
\label{eq:tct3}
\bar{f}_{(2)}^{ij} &=& \Lambda_{k}^{i} \Lambda_{l}^{j} f_{(2)}^{kl}, \\
&\vdots& \nonumber
\end{eqnarray}
Thus the action of an arbitrary unitary operation preserves the form of $\hat{\rho}= e^{\hat{f}(\hat{q},\hat{p})}$ by only evolving the coefficients of $f$  (coefficients get multiplied by the $\Lambda_i^j$ effecting the transformation of the symmetric products which in turn depend only on the first order/linear transformations).

Now, consider a factorizable two-mode state,
\begin{eqnarray}
\label{eq:facstate}
\hat{\rho}_a \hat{\rho}_b = e^{ \hat{f}(\hat{q}_a, \hat{p}_a) + \hat{g}(\hat{q}_b, \hat{p}_b) }.
\end{eqnarray}
The exponent of the two-mode density operator $\hat{\rho}_{ab} = \hat{\rho}_a \hat{\rho}_b$ is given by $\hat{h} = \hat{f} + \hat{g}$ and can be expanded in terms of $\vec{\xi} = (\hat{q}_a,\hat{p}_a,\hat{q}_b,\hat{p}_b)$ as
\begin{eqnarray}
\label{eq:twopse}
\hat{h} = h_{(0)} + h_{(1)}^i \hat{\xi}_i + h_{(2)}^{ij} \{ \hat{\xi}_i \, \hat{\xi}_j \}_s + \cdots,
\end{eqnarray}
where the coefficients $h$ results from those of $f$ and $g$, that is,
\begin{eqnarray}
\label{eq:vancf}
h_{(0)} &=& f_{(0)} + g_{(0)} \nonumber \\
h_{(1)}^i &=& \left\{
\begin{array}{lr}
f_{(1)}^i & \mbox{if $i \le 2$} \\
g_{(1)}^{i-2} & \mbox{if $i > 2$}
\end{array}
\right. \nonumber \\
h_{(2)}^{i_1 i_2} &=& \left\{
\begin{array}{lr}
f_{(2)}^{i_1 i_2} & \mbox{if $i_1,i_2 \le 2$} \\
g_{(2)}^{i_1-2 \, \,i_2-2} & \mbox{if $i_1,i_2 > 2$} \\
0 & \mbox{otherwise}
\end{array}
\right. \nonumber \\
h_{(n)}^{i_1 i_2 \cdots i_n} &=& \left\{
\begin{array}{lr}
f_{(n)}^{i_1 i_2 \cdots i_n} & \mbox{if $i_k \le 2, \forall k$} \\
g_{(n)}^{i_1-2 \, \, i_2-2 \, \, \cdots \, \, i_n-2} & \mbox{if $i_k > 2, \forall k$} \\
0 & \mbox{otherwise}
\end{array}
\right.
\end{eqnarray}
It is clear that the factorizable state is conditioned by the vanishing coefficients of the mixed moments such as $h_{(2)}^{13}$ and $h_{(2)}^{14}$. Note that the mixed moments appear from second order onward. For the moments higher than first order, all the coefficients $h_{(n)}^{i_1 i_2 \cdots i_n}$ are of mixed moments except the cases where $\forall i_k \le 2$ and $\forall i_k > 2$.

The transformations by the beam splitter of the coefficient tensors $h_{(n)}$ are given by Eq.~(\ref{firstorderBS}), similarly to Eqs.~(\ref{relation})-(\ref{eq:tct3}). The first order coefficients are transformed to
\begin{eqnarray}
\label{eq:firmomtrans}
\bar{h}_{(1)}^{i} = \Lambda^i_j h_{(1)}^{j}
\end{eqnarray}
where
\begin{eqnarray}
\Lambda =
\begin{pmatrix}
t & 0 & -r & 0 \\
0 & t & 0 & -r \\
r & 0 & t & 0 \\
0 & r & 0 & t
\end{pmatrix}.
\end{eqnarray}
In general, the transformations of the $n$-th order coefficients $h_{(n)}$ are given as
\begin{eqnarray}
\bar{h}_{(n)}^{i_1i_2 \cdots i_n} = \Lambda^{i_1}_{j_1}\Lambda^{i_2}_{j_2} \cdots \Lambda^{i_n}_{j_n} h_{(n)}^{j_1j_2 \cdots j_n}.
\end{eqnarray}

In order for output to be factorizable/uncorrelated, all the vanishing coefficients in Eq.~(\ref{eq:vancf}) must be retained after the transformation. For factorizable output we have no restriction on the zero-th and first order moments. For the second order moments, we have the conditions,
\begin{eqnarray}
\label{eq:faccon2m}
\bar{h}_{(2)}^{i\, \, j+2} = 0,
\end{eqnarray}
for each $i,j \le 2$. The other conditions $\bar{h}_{(2)}^{i+2\, \, j}=0$ are redundant due to the symmetry of coefficient tensor $\bar{h}_{(2)}$. The conditions (\ref{eq:faccon2m}) are rewritten in terms of the transformation matrix $\Lambda$ and the untransformed coefficients $h_{(2)}$ as
\begin{eqnarray}
\Lambda^i_k \Lambda^{j+2}_l h_{(2)}^{kl} = 0.
\end{eqnarray}
For $i \le 2$, $\Lambda^i_i = t$, $\Lambda^i_{i+2} = -r$, and $\Lambda^i_{k} = 0$ if $k \ne i$ or $i+2$. Similarly, for $j \le 2$, $\Lambda^{j+2}_j = r$, $\Lambda^{j+2}_{j+2} = t$, and $\Lambda^{j+2}_{k} = 0$ if $k \ne j$ or $j+2$. Further $h_{(2)}^{kl} \ne 0$ only if both $k,l \le 2$ or both $k,l > 2$. When these are applied, the conditions become
\begin{eqnarray}
\label{eq:facconsm}
r \, t \left( h_{(2)}^{ij} - h_{(2)}^{i+2\, \, j+2} \right) = 0.
\end{eqnarray}
The case of $r = 0$ or $t = 0$ is irrelevant as it implies either simply a mirror or no beamsplitter, and we consider the case of $t \ne 0$ and $r \ne 0$. Noting that $h_{(2)}^{ij} = f_{(2)}^{ij}$ and $h_{(2)}^{i+2 \, \, j+2} = g_{(2)}^{ij}$ for $i,j \le 2$ (when the inputs are factorizable), the conditions imply that
\begin{eqnarray}
\label{eq:facconsmm}
f_{(2)}^{ij} = g_{(2)}^{ij}.
\end{eqnarray}
This is the condition for the output to be factorizable for factorizable input states that are Gaussian (i.e. second order). As we shall now see in fact it is only second order input states that are permitted if the output is to be uncorrelated.

Similar arguments can be applied to the higher order coefficient tensors $h_{(n)}$. The factorizable conditions, similar to Eq.~(\ref{eq:facconsm}), are given as
\begin{eqnarray}
\label{eq:facconhm}
 r^j t^{n-j} f_{(n)}^{i_1 i_2 \cdots i_n} + (-r)^{n-j} t^j g_{(n)}^{i_1 i_2 \cdots i_n} = 0,
\end{eqnarray}
for $j=1,2,\ldots, n-1$ and all $i_k$. Assuming $t \ne 0$ and $r \ne 0$, the conditions are reduced for odd $n$-th order moments into
\begin{eqnarray}
t f_{(n)}^{i_1 i_2 \cdots i_n} \pm r g_{(n)}^{i_1 i_2 \cdots i_n} &=& 0, \nonumber \\
r f_{(n)}^{i_1 i_2 \cdots i_n} \mp t g_{(n)}^{i_1 i_2 \cdots i_n} &=& 0,
\end{eqnarray}
which clearly result in, for all $i_k$,
\begin{eqnarray}
\label{eq:faccongm}
f_{(n)}^{i_1 i_2 \cdots i_n} = g_{(n)}^{i_1 i_2 \cdots i_n} = 0.
\end{eqnarray}
Similarly, for even $n$-th order moments, the conditions in Eq.(\ref{eq:facconhm}) are reduced to
\begin{eqnarray}
t^2 f_{(n)}^{i_1 i_2 \cdots i_n} &\mp& r^2 g_{(n)}^{i_1 i_2 \cdots i_n} = 0, \nonumber \\
f_{(n)}^{i_1 i_2 \cdots i_n} &\pm& g_{(n)}^{i_1 i_2 \cdots i_n} = 0, \nonumber \\
r^2 f_{(n)}^{i_1 i_2 \cdots i_n} &\mp& t^2 g_{(n)}^{i_1 i_2 \cdots i_n} = 0,
\end{eqnarray}
also leading to the conditions (\ref{eq:faccongm}). We have the conditions (\ref{eq:faccongm}) hold for all moments higher than second order.

All terms above second order must be zero in an input state in order for the output from a beam splitter to be factorizable. Furthermore the conditions in Eq.~(\ref{eq:facconsmm}) specify that the two states should have the same second order coefficients whereas the first order terms are not restricted, i.e. Gaussian inputs differing only by a displacement give factorisable output. 

Let us examine why they may differ by a displacement.
Arbitrary displacements of $n$-th order input states creates additional terms of order $n-1$ and lower which may then be subsumed into the coefficients of existing terms. Considering the restriction to second order the displacements may only introduce linear terms and these terms cannot alter the factorisability of the output (these linear terms when transformed, according to Eq.~(\ref{firstorderBS}), remain linear). 

Alternatively, we may see this condition in the Wigner representation. Based on the neccessity of cross terms canceling in a general form of the Wigner function, we can displace identical Gaussian input states each by arbitrary amounts and still retain factorizability of the output. A single-mode Wigner function is given by \cite{Leonhardt}
\begin{eqnarray}
W(q,p)=\frac{1}{2 \pi} \int_{-\infty}^\infty e^{i p y} \bra{q- \frac{y}{2}} \hat{\rho} \ket{q+ \frac{y}{2}} dy
\end{eqnarray}
where $|q \rangle$ are eigenstates of a quadrature variable $\hat{q}$.
The Wigner function for a general Gaussian state is the exponential of a quadratic function of $z=\frac{1}{\sqrt{2}}(q+ i p)$. One form, of Argawal \cite{Argawal1989}, that lends itself to describing many states is
\begin{eqnarray}
& & W_1(z)=\frac{1}{\pi \sqrt{\tau^2-4 |\mu|^2}} \times \nonumber\\
& & \exp \left( \frac{- \left[ \mu (z-z_0)^2 + \mu^* (z^* - z_0^*)^2 + \tau |z- z_0|^2 \right]}{\tau^2-4 |\mu|^2} \right) \nonumber\\
\label{WignerGaussian} \end{eqnarray}
which corresponds to the density matrix
\begin{eqnarray}
\label{eq:ggsief}
\hat{\rho} &=& \frac{2}{\sqrt{e^{2\phi}-1}} \exp \left(-2 e^{-\phi} \textrm{arcosh}( \coth(\phi)) \left[ \mu (\hat{a}-z_0)^2 \right. \right. \nonumber \\
& & \left. \left. + \mu^* (\hat{a}^\dag - z_0^*)^2 + \tau(\hat{a}-z_0)(\hat{a}^\dag - z_0^*)\right] \right)
\end{eqnarray}
with
\begin{eqnarray}
\phi  = \frac{1}{2} \ln( 4(\tau^2 - 4 |\mu|^2) ),&& \\
\langle \hat{a} \rangle = z_0, \, \, \langle \hat{a}^\dag \rangle = z_0^*,&& \\
 \Delta \hat{a} = -2 \mu^*, \, \, \Delta \hat{a}^\dag= -2 \mu,&& \\
\langle \hat{a}^\dag \hat{a} \rangle = \tau - \frac{1}{2} + |z_0|^2.&&
\end{eqnarray}
The state is entirely specified by $\mu$, $\tau$ and $z_0$  where $z_0$ describes the degree of displacement from the vacuum state. 
The requirement that a density matrix corresponds to a legitimate physical state (i.e. is positive semidefinite) is expressed as
\begin{eqnarray}
\sqrt{\tau^2 - 4 |\mu|^2} \ge \frac{1}{2} \label{WignerPositiveSDef}
\end{eqnarray}
in terms of the parameters $\tau$ and $\mu$ \cite{Argawal1988}.

The state described by (\ref{WignerGaussian}) is actually quite a general form; it is a mixed state (unless there is equality in criterion (\ref{WignerPositiveSDef})) accounting for noisy processes (when $\mu \ne 0$). It was proposed to provide a description of interferometers with losses \cite{Argawal1987} and the density matrix given by (\ref{WignerGaussian}) represents a wide variety of states formed by non-linear optics experiments \cite{Argawal1988}.

We are concerned with inputs of the form $\hat{\rho}=e^{f_{(2)}(\hat{q}_a, \hat{p}_a)} e^{g_{(2)}(\hat{q}_b, \hat{p}_b)}$, such factorizable two-mode states are defined by the product of the Wigner functions describing each mode. For the output after the beamsplitter to retain factorizability requires that the output is in the form of the product of Wigner functions describing each mode, i.e.
\begin{eqnarray}
W_a(q_a, p_a) W_b(q_b, p_b) \underset{BS}{\rightarrow} W'_a(q_a, p_a) W'_b(q_b, p_b).
\end{eqnarray}
Since we are writing the Gaussian Wigner functions in an exponential form, factorizability is again equivalent to there being no cross terms in the exponent of the output---the Wigner function of the output is of the form $e^{f'_{(2)}(\hat{q}_a, \hat{p}_a)} e^{g'_{(2)}(\hat{q}_b, \hat{p}_b)}$.

Thus impinging a Gaussian state of the form (\ref{WignerGaussian}) into one input port of a beam splitter and another Gaussian into the other mode (with corresponding parameters $\mu'$ and $\tau'$) the output is an exponential and we find that the coefficients of the cross terms $q_a q_b$, $p_a p_b$, $q_a p_b$ are
\begin{eqnarray}
\frac{2 r t (\mu' + \mu'^* + \tau')}{\tau'^2 - 4 |\mu'|^2} &-& \frac{2 r t (\mu + \mu^* + \tau)}{\tau^2 - 4 |\mu|^2}, \\
\frac{2 r t (-\mu' - \mu'^* + \tau')}{\tau'^2 - 4 |\mu'|^2} &-& \frac{2 r t (-\mu + -\mu^* + \tau)}{\tau^2 - 4 |\mu|^2}, \\
\frac{2 i r t (\mu' - \mu'^*)}{\tau'^2 - 4 |\mu|^2} &-& \frac{2 i r t (\mu + -\mu^*)}{\tau^2 - 4 |\mu|^2},
\end{eqnarray}
respectively (the coefficient of the other cross term $p_a q_b$ is identical to that of $q_a p_b$). Setting these equal to zero yields two solutions: $(\mu=0, \tau=0)$ and $(\mu=\mu', \tau=\tau')$; the first is rejected as it is unphysical (it fails the positive semidefinite criterion Eq.~(\ref{WignerPositiveSDef}); the second is valid. Note that it places no restriction on the value of $z_0$, thus when considering Gaussian input fields an arbitrary displacement has no effect on the factorizability of the outputs.

\section{Summary}
We discussed the role of correlated states in quantum state engineering, in particular in conditional measurement schemes. It was highlighted that the correlations required to implement such a scheme need not be entanglement of the modes but rather that classical correlations suffice. This led us to consider what class of states input to a beam splitter exit as factorizable states; this class of states possess neither classical correlations nor quantum entanglement. By considering the action of the beam splitter on arbitrary inputs we found that only in the case that the two input states are identical Gaussians, apart from an arbitrary displacement applied to each mode, do we obtain an output that is factorizable.

\acknowledgements
We acknowledge financial support from Northern Ireland's Department for Employment and Learning, the Korea Research Foundation Grant funded by the Korean Government (KRF-2008-313-C00188), Ente Cassa di Risparmio di Firenze and CNR-RSTL and EPSRC.

\end{document}